\documentclass[aps,prb,twocolumn,showpacs,groupedaddress,preprintnumbers,amsmath,amssymb,raggedbottom]{revtex4}
\usepackage{amsmath,amssymb,epsfig,color,pstricks,bm,alltt,graphicx}

\usepackage[colorlinks,plainpages=false,linkcolor=black,urlcolor=blue,citecolor=black,pdfpagemode=UseNone,pdfstartview=FitBH]{hyperref}

\definecolor{MyDarkBlue}{rgb}{0,  0.3,  0.9}
\definecolor{MyDarkBlack}{rgb}{0,  0,  0}
\newcommand \modified[1]{\textcolor{MyDarkBlack}{#1}}

\definecolor{darkgreen}{rgb}{0,0.5,0}

\begin{document}

\title{Electronic structure of Pr$_{2-x}$Ce$_{x}$CuO$_4$ studied
via ARPES and LDA+DMFT+$\Sigma_{\textbf{k}}$}
\author {I.\,A.~Nekrasov}
\author {N.\,S.~Pavlov}
\author {E.\,Z.~Kuchinskii}
\author {M.\,V.~Sadovskii}
\affiliation {Institute for Electrophysics, Russian Academy of Sciences,
Ekaterinburg 620016, Russia}

\author {Z.\,   V. Pchelkina}
\affiliation {Institute for Metal Physics, Russian Academy of Sciences,
Ekaterinburg 620219, Russia}

\author{V.~B.~Zabolotnyy}
\affiliation{Institute for Solid State Research, IFW-Dresden, P.O.Box 270116, D-01171 Dresden, Germany}
\author{J.~Geck}
\author{B.~B\"{u}chner}
\author{S.\,V.~Borisenko}
\affiliation{Institute for Solid State Research, IFW-Dresden, P.O.Box 270116, D-01171 Dresden, Germany}

\author{D.~S.~Inosov}
\affiliation{Max-Planck-Institute for Solid State Research, Heisenbergstrasse 1, D-70569
Stuttgart, Germany}
\affiliation{Institute for Solid State Research, IFW-Dresden, P.O.Box 270116, D-01171 Dresden, Germany}

\author{A.~A.~Kordyuk}
\affiliation{Institute for Solid State Research, IFW-Dresden, P.O.Box 270116, D-01171 Dresden, Germany}
\affiliation{Institute of Metal Physics of National Academy of Sciences of Ukraine, 03142 Kyiv,  Ukraine}

\author{M.~Lambacher}
\author{A.~Erb}
\address{Walther-Mei\ss ner-Institut, Bayerische Akademie der Wissenschaften,
 Walther-Mei\ss ner Strasse 8, 85748 Garching, Germany}

\date{\today}

\begin{abstract}

The electron-doped Pr$_{2-x}$Ce$_{x}$CuO$_4$ (PCCO) compound in the pseudogap
regime ($x\approx0.15$) was investigated using  angle-resolved photoemission spectroscopy (ARPES) and the
generalized dynamical
mean-field theory (DMFT) with the {\bf k}-dependent self-energy (LDA+DMFT+$\Sigma_{\textbf{k}}$).
Model parameters (hopping integral values and local Coulomb interaction strength) for the effective
one-band Hubbard model were calculated by the local density approximation (LDA) with numerical
renormalization group method (NRG) employed as an ``impurity solver'' in
DMFT computations.  An ``external'' {\bf k}-dependent self-energy $\Sigma_{\textbf{k}}$ was used to
describe interaction of correlated conducting electrons with short-range
antiferromagnetic (AFM) pseudogap fluctuations.
Both experimental and theoretical spectral functions and Fermi surfaces (FS)
were obtained and compared demonstrating  good semiquantitative agreement.
For both experiment and theory normal state spectra of nearly optimally doped  PCCO show clear evidence
for a pseudogap state with AFM-like nature.
Namely, folding of quasiparticle bands as well as  presence of the ``hot spots'' and ``Fermi arcs'' were observed.
\end{abstract}


\maketitle

\section{Introduction}

Many experimental and theoretical papers have been seeking
to \modified{describe} the nature of high-temperature superconductivity (HTSC)
in cuprates. In contrast to  the normal (Fermi-liquid) metal, HTSC compounds show
many abnormal properties for temperatures above the superconducting transition\modified{; the normal state pseudogap being a notorious example}\cite{pseudogap}.
\modified{The origin of this anomalous state is usually attributed either to superconducting fluctuations (precursor Cooper pairing) \cite{fluctuations} or to some order parameter competing with superconductivity \cite{Bi2212, nonmonPG}, e.g. AFM fluctuations, incommensurate or fluctuating charge density waves (CDW), stripes, etc.}

Recently a generalized LDA+DMFT+$\Sigma_{\textbf{k}}$ computational scheme
was proposed to describe the pseudogap state in strongly correlated
systems, by accounting for nonlocal AFM (or CDW) fluctuations with short-range order \cite{fsdistr,dmftsk,opt}. Its relation to other theoretical
DMFT-like\cite{DMFT_method} approaches to the pseudogap state was discussed
e.g. in Ref.~\onlinecite{Bi2212}.
Both model computations and those for
real systems were done \cite{Bi2212,NCCO_work}. This approach, for instance, allowed \modified{one to describe}
 the experimentally observed partial Fermi surface
(FS) ``destruction''\cite{fsdistr}, which was theoretically studied for
\modified{hole-doped} HTSC prototype system Bi$_2$Sr$_2$CaCu$_2$O$_{8-\delta}$
(Bi2212)\cite{Bi2212} and \modified{electron-doped} one Nd$_{1.85}$Ce$_{0.15}$CuO$_4$ (NCCO)\cite{NCCO_work}.
Two-particle properties can also be described \modified{by} this approach\cite{opt},
e.g. calculated optical spectra in the pseudogap state
\modified{compare} well with experimental data for Bi2212\cite{Bi2212} and NCCO\cite{NCCO_work}.

In this paper we study the electron-doped Pr$_{2-x}$Ce$_{x}$CuO$_4$ (PCCO)
in the pseudogap state ($x=0.15$) using
the generalized LDA+DMFT+$\Sigma_{\textbf{k}}$ computational scheme
\cite{fsdistr,dmftsk,opt} and ARPES measurements\cite{ARPES, ARPES1}.
We present here both experimental and theoretical quasiparticle spectral
functions and  Fermi surfaces. These are found to agree well with each
other supporting competing order parameter fluctuations as origin of
the pseudogap instead of superconducting scenario.

\section{Computational details}

The crystal structure\cite{Crystal_structure} of Pr$_2$CuO$_4$ has tetragonal symmetry and the space group is I4/mmm.
Lattice constants\cite{Crystal_structure} are \modified{a=b=3.962, c=12.154 {\AA}.}
There are two crystallographic types of oxygen atoms in Pr$_2$CuO$_4$:
the first one belongs to CuO$_2$ layer and the second  is located within the Pr atoms.
The atomic positions in the elementary cell are as follows:
Cu -- 2a(0,0,0), \hspace{4 em} O1 -- 4c(0,0.5,0), 
Pr -- 4e(0,0,0.35171), O2 -- 4d(0,0.5,0.25).\cite{Crystal_structure}

\begin{figure}[h]
\includegraphics[clip=true,scale=0.25]{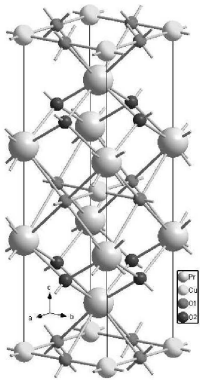}
\caption{The crystal structure of Pr$_2$CuO$_4$.
A middle size grey spheres correspond to the copper atoms,
a small dark and black spheres are  O1 and O2 atoms, respectively
and a big grey spheres --- praseodymium atoms.
\label{Crystal_Structure_PrCuO}
}
\end{figure}

In Fig.~\ref{Crystal_Structure_PrCuO} the crystal structure of Pr$_2$CuO$_4$ is shown.
\modified{Middle} size grey spheres correspond to the copper atoms,  \modified{little} dark and black spheres represent
O1 and O2 atoms, big grey spheres show praseodymium positions.
Clearly visible quasi two-dimensional nature of these compound\modified{s}
determines its physical properties.
Physically most interesting are the CuO$_2$ layers. Those layers provide
antibonding Cu-3$d$($x^2-y^2$) partially filled orbital, whose
dispersion crosses the Fermi level.
Thus we are using this effective Cu-3$d$($x^2-y^2$) antibonding band as a ``bare''
band in DMFT computations.

\begin{figure}[!b]

\includegraphics[width=0.85\columnwidth]{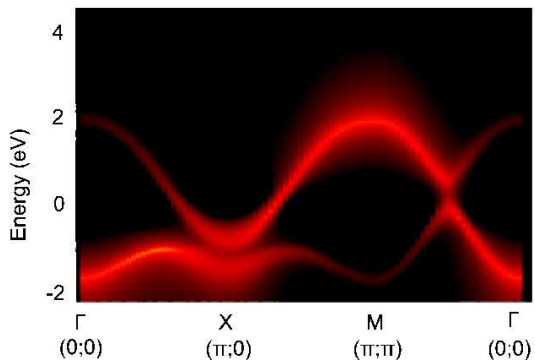}
\caption{LDA+DMFT+$\Sigma_k$ quasiparticle energy dispersion of PCCO Cu-3$d$($x^2-y^2$) orbital
for high symmetry directions of square Brillouin zone. \modified{The Fermi level is zero.}
\label{Contour_plot}}
\end{figure}

Electronic structure of PCCO was investigated within generalized LDA+DMFT+$\Sigma_{\textbf{k}}$
scheme.\cite{fsdistr,dmftsk,opt}
This scheme incorporates the density functional theory in local density approximation (LDA)\cite{LDA_method}
and the dynamic mean-field theory (DMFT)\cite{DMFT_method} with ``external'' momentum-dependent self-energy
$\Sigma_{\textbf{k}}$\cite{dmftsk}.

As a first step the LDA band structure calculation was performed.
Using crystal structure data, the electronic band structure was obtained with the linearized
muffin-tin orbitals (LMTO) method\cite{LMTO}.
Further hopping integral values were calculated for effective Cu-3$d$($x^2-y^2$) Wannier function
within the \modified{$N$}-th order LMTO (NMTO) framework.\cite{NMTO}
Corresponding hopping integral values are $t=-0.4385$, $t'=0.1562$, $t''=0.0976$.
The value of Coulomb interaction on effective Cu-3$d$($x^2-y^2$) orbital $U$=1.1~eV
was obtained via constrained LDA computations\cite{Gunnarsson}.
Secondly, the DMFT calculations \cite{DMFT_method},
which take the hopping integrals and the Coulomb interaction as input parameters, were performed.

To account for the AFM spin fluctuations, \modified{a} two-dimensional model of \modified{the} pseudogap state is applied.\cite{AFM_fluctuations}
Corresponding \textbf{k}-dependent self-energy $\Sigma_{\textbf{k}}$\cite{pseudogap,AFM_fluctuations}
describes nonlocal correlations induced by (quasi) static short-range collective
Heisenberg-like AFM spin fluctuations.
The quasi static approximation for AFM fluctuations necessarily limits
our approach to high - enough temperatures (energies not so close to the Fermi
level)\cite{AFM_fluctuations}, so that in fact we are unable to judge e.g. on 
the nature of low temperature (energy) damping in our model.
Thus we avoid here possible discussion of whether the damping corresponds to marginal
or regular Fermi liquid. Moreover as shown in Ref.~\onlinecite{dmftsk} for the
``hot-spot'' corresponding self-energy has essentially non Fermi liquid behavior.

The $\Sigma_{\textbf{k}}$ definition contains two important parameters:
the pseudogap energy scale (amplitude) $\Delta$, representing the energy scale
of fluctuating \modified{SDW}, and the spatial correlation
length $\xi$.
\modified{The latter} is usually determined from experiment.
The $\Delta$ value was calculated as described in Ref. \onlinecite{dmftsk}
and found to be 0.275~eV.
The value of correlation length was taken to be 50 lattice constants, in accordance with the typical
value obtained in neutron scattering experiments on NCCO \cite{NCCOxi}.
To solve DMFT equations numerical renormalization group (NRG\cite{NRG,BPH}) was
employed as an ``impurity solver''.
Corresponding temperature of DMFT(NRG) computations was 0.011~eV and
electron concentration was n=1.145.

\begin{figure}
\includegraphics[clip=true,width=1.\columnwidth]{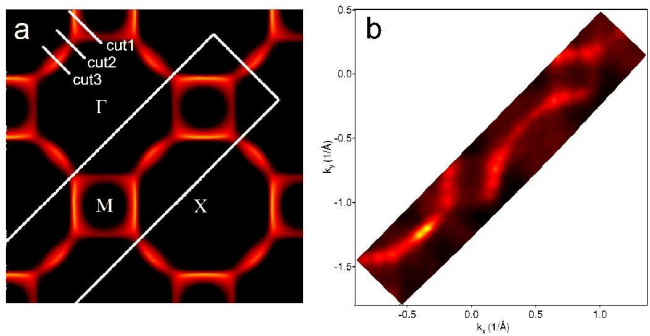}\\
\caption{ (a) Extended Fermi surfaces for PCCO --- LDA+DMFT+$\Sigma_{\textbf{k}}$ data.
White rectangle on panel (a) schematically shows the part of reciprocal space
measured experimentally (panel b). Lower left corner is X-point ($\pi,0$).
\label{FermiSurface}}
\end{figure}

\section{Experimental details}
Photoemission  experiments were performed at UE112-PGM beamline at BESSY using SCIENTA SES100 analyzer.
Typical energy and angular resolution for the excitation energy ($h\nu=100$\;eV) used in this study
were 20\;meV and $0.2^\circ$ respectively.
Samples of Pr$_{1.85}$Ce$_{0.15}$CuO$_{4+\delta}$ were grown using traveling solvent floating zone
technique and annealed to achieve optimal $T_\textup{c}$ of 25K with transition width of 1K, 
\modified{which resulted from improved growth conditions \cite{Lambacher}.
Similarly the with (FWHM) of X-ray rocking curves was less than 0.08$^\circ$, signaling high quality of the samples.}

For the photoemission measurements  the samples were mounted on a cryomanipulator and
cleaved in situ in \modified{ultra high vacuum} with a base pressure $\lesssim 1\cdot10^{-10}$\;mBar. During the whole experiment,
including the temperature cycling, when the sample was heated to room temperature
and then cooled back again, no observable aging effects were detected.

\section{Results and discussion}

Generally speaking,  finite temperature and interaction lead to notable life-time effects.
Thus, instead of quasiparticle dispersions obtained in usual band structure calculations one has to deal with
corresponding spectral function  $A(\omega,\textbf{k})$:
\begin{equation}
A(\omega,\textbf{k})=-\frac{1}{\pi}{\rm Im} G(\omega,\textbf{k}),
\label{specf}
\end{equation}
where $G(\omega,\textbf{k})$ is the retarded Green's function obtained via LDA+DMFT+$\Sigma_{\textbf{k}}$
scheme\cite{fsdistr,dmftsk,opt}.

Color plots, whose intensity encodes the function values became a traditional
and convenient way of representing these
multiple variable functions. Such a color plot of LDA+DMFT+$\Sigma_{\textbf{k}}$ quasiparticle spectral function
(\ref{specf}) of copper
3$d$($x^2-y^2$) orbital is presented in Fig.~\ref{Contour_plot}.
\modified{Width} of \modified{the} quasiparticle spectral  function in the color plot is inversely proportional to the quasiparticle life time.
The calculated quasiparticle band dispersion has minimum at $\Gamma$ point (-1.52~eV) and maximum at M point (2~eV).
It crosses the Fermi level along X-M as well as M-$\Gamma$ directions.
Because of AFM pseudogap fluctuations there is \modified{a} well detectable (quasi) folding of the
quasiparticle band, reflected in the formation of
the so called ``shadow'' band, which has its maxima at $\Gamma$-point and minima at M-point.
However, because of the short-range nature of the antiferromagnetic order, this does not result in a complete folding,
as it would be the case for
a long-range AFM order. Namely, quasiparticle band and the  ``shadow'' band are not exactly the same.
No real band gap opens at ($\pi/2,\pi/2$) point.  Nevertheless,  suppression of  the spectral weight is clearly
detectable in the vicinity of
X $(\pi, 0)$ point, thus  signaling  opening of the pseudogap, which in this case can be viewed as a precursor of
the real band gap. Splitting takes place between the quasiparticle band and the ``shadow'' band with the value \modified{of} about
2$\Delta$.

\begin{figure}[!b]
\includegraphics[clip=true,width=1\columnwidth]{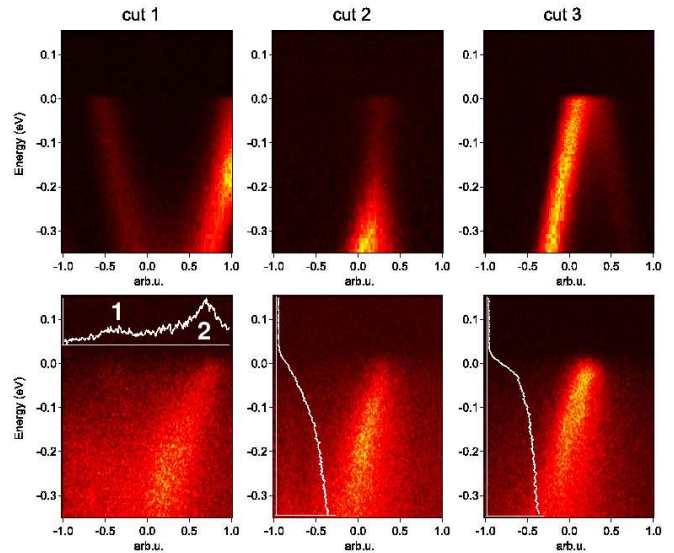}
\caption{Energy--momentum intensity distributions for the specific cuts drawn in Fig.~\ref{FermiSurface}
(upper panels --- theoretical data, lower panels --- experimental photoemission intensity).
To judge about the absolute
intensities of the ``shadow'' (1) and main band (2) cut 1 contains an
MDC curve integrated in an energy window 60 meV
centered at the Fermi level (FL). Similarly integral EDC for cut 2 (``hot 
spot'') shows suppression of the intensity at the
FL as compared to cut 3, which is located further away from
the ``hot spot''.
The FL is zero.\label{Cuts}}
\end{figure}

In Fig.~\ref{FermiSurface} an extended picture of PCCO Fermi surfaces is presented
(panel (a) --- LDA+DMFT+$\Sigma_{\textbf{k}}$ results, panel (b) --- experimental ARPES data).
Strictly speaking Fig.~\ref{FermiSurface} is \modified{a} color map in reciprocal space of \modified{the} corresponding
spectral function plotted \modified{at  the} Fermi level.
FS is clearly visible as reminiscence of non-interacting band close to the first Brillouin zone border and around
\modified{$(\pi/2,\pi/2)$} point (so called Fermi arc), where the quasiparticle band crosses the Fermi level.
There is \modified{an} interesting physical effect of  partial ``destruction'' of the FS
observed in the ``hot spots'', points that are located at  the intersection of the FS and its AFM umklapp replica.
This FS ``destruction''  occurs because of the  strong electron scattering on the antiferromagnetic
(AFM) spin (pseudogap) fluctuations on the copper atoms.
Also the ``shadow'' FS is visible as it should be for AFM folding.  \modified{As no long-range order is present in the
underdoped phase} the ``shadow'' FS has weaker intensity with respect to FS.
Another evidence of presence of electron pockets can be seen in the
experimental FS map shown in Fig~\ref{FermiSurface}b. One pocket is centered at point with
coordinates (0,0.8$\pi$) and the other at (0.8$\pi$,0).
Again, in agreement with the theoretical prediction, the pocket
``sides'' which coincide with originally unreconstructed FS have higher
intensity, while the newly appearing ``replicas/shadows'' have weaker intensity.

 The PCCO FS is very similar to that of 
 Nd$_{2-x}$Ce$_{x}$CuO$_4$ (NCCO), which belongs to the same family of
supperconductors. The NCCO was recently studied both theoretically\cite{NCCO_work} and experimentally
\cite{Armitage}.

Let us compare theoretical (upper line) and experimental (lower line) energy quasiparticle dispersion
for most characteristic cuts introduced in Fig.~\ref{FermiSurface} (see Fig.~\ref{Cuts}).
Theoretical data \modified{were} multiplied by the Fermi function \modified{at a temperature of 30K} and convoluted
with \modified{a} Gaussian to simulate the effects of experimental resolution, with further artificial noise added.

The Cut 1 intersects quasiparticle and ``shadow'' Fermi surfaces close to the \modified{Brillouin zone} border.
One can find here a ``fork''-like structure formed by the damped ``shadow'' band (-0.5-0 \modified{arb. u.}) and
better defined quasiparticle band (0.5-1 a.u.). This structure corresponds to preformation of
FS cylinder around ($\pi$,0) point.
The Cut 2 goes exactly through the ``hot spot''.
Here we see \modified{a} strong suppression of the quasiparticle band around the Fermi level as it is also shown
in Fig.~\ref{Contour_plot}. The Cut 3 crosses the Fermi arc, where  we can see a very well defined
quasiparticle band. However \modified{weak} intensity ``shadow'' band is also present. For the
case of long range AFM order and complete folding of electronic structure, FS and its ``shadow''
should form a closed FS sheet
around ($\pi/2$, $\pi/2$) point, while in the current case the part of the pocket formed by the ``shadow'' band is not
so well defined in momentum space.
As can be seen there is \modified{a} good correspondence between the calculated and experimental data in terms of the  above
described behavior, which is also similar
to the results reported for Nd$_{2-x}$Ce$_{x}$CuO$_4$ (NCCO) in our earlier work.\cite{NCCO_work}
\section{Conclusion}
In this work the LDA+DMFT+$\Sigma_{\textbf{k}}$ was performed for electron-doped Pr$_{2-x}$Ce$_{x}$CuO$_4$ compound
in the pseudogap regime. The LDA+DMFT+$\Sigma_{\textbf{k}}$ calculation shows that Fermi-liquid behavior is still
conserved far away from the ``hot-spots'', while the  destruction of the Fermi surface observed in the vicinity of
``hot spots'' is due to strong scattering of correlated electrons
on short-range antiferromagnetic (pseudogap) fluctuations. Comparison between experimental ARPES and
LDA+DMFT+$\Sigma_{\textbf{k}}$ data  reveals a
good semiquantitative agreement. The experimental and theoretical results obtained once again support the
AFM scenario of pseudogap formation not only in hole doped HTSC systems\cite{Bi2212} but also
in electron doped ones~\cite{NCCO_work}.

We thank Thomas Pruschke for providing us the NRG code. This work is supported by
RFBR grants 08-02-00021, 08-02-91200 and RAS programs ``Quantum physics of condensed matter'' 
and ``Strongly correlated electrons in solids''.
IN and ZP are supported by 
Grants of President of Russia MK-3227.2008.2(ZP) and MK-614.2009.2(IN),
Russian Science Support Foundation (IN) and
\modified{the} Dynasty Foundation (ZP).
The experimental measurements for this study \modified{were} possible owing to the
financial support of  Forschergruppe FOR538 and by the DFG under Grant No. KN393/4.


\end{document}